\documentclass[a4paper,11pt]{article}
\usepackage{jheppub} 
\usepackage{orcidlink}
\usepackage{microtype}

\usepackage{mathtools}
\usepackage{amsmath}
\usepackage{amsfonts}
\usepackage{amssymb}
\usepackage{bbold}
\usepackage{bm}
\usepackage{bbm}
\usepackage{esint}
\usepackage{booktabs}

\newcommand{\bH}{{\bf H}}
\newcommand{\bB}{{\bf B}}
\newcommand{\bL}{{\bf L}}
\newcommand{\bJ}{{\bf J}}
\newcommand{\bP}{{\bf P}}
\newcommand{\bK}{{\bf K}}
\newcommand{\bG}{{\bf G}}
\newcommand{\bR}{{\bf R}}
\newcommand{\br}{{\bf r}}

\newcommand{\bS}{{\bf S}}
\newcommand{\bk}{{\bf k}}
\newcommand{\bv}{{\bf v}}
\newcommand{\bp}{{\bf p}}

\newcommand{\bsigma}{{\bm\sigma}}

\newcommand{\sH}{{\sf H}}
\newcommand{\sM}{{\sf M}}

\newcommand{\bpartial}{\boldsymbol{\partial}}

\long\def\exclude#1{}

\newcommand{\GF}{G_{\rm F}}

\title{Single-wave solutions of the neutrino fast flavor system. Part~I.~Mechanical properties.}

\author[a,b,c]{Damiano F.\ G.\ Fiorillo \orcidlink{0000-0003-4927-9850}} 
\affiliation[a]{Deutsches Elektronen-Synchrotron DESY,
Platanenallee 6, 15738 Zeuthen, Germany}
\affiliation[b]{Istituto Nazionale di Fisica Nucleare (INFN), Sezione di Napoli,
Complesso Universitario di Monte Sant’Angelo, Via Cintia, 80126 Napoli, Italy}
\affiliation[c]{Gran Sasso Science Institute (GSSI), L’Aquila, Italy}

\author[d]{and Georg G.\ Raffelt
\orcidlink{0000-0002-0199-9560}}
\affiliation[d]{Max-Planck-Institut f\"ur Physik, Boltzmannstr.~8, 85748 Garching, Germany}

\abstract{A dense neutrino plasma can exhibit collective flavor evolution caused by neutrino--neutrino refraction. Recently, a new class of exact nonlinear inhomogeneous solutions was discovered: single-wave (SW) solutions of the fast flavor system. The key property is that the flavor occupation numbers remain homogeneous, whereas the field of flavor coherence varies spatially with a single wave vector. The equations of motion for this structure resemble those of a collection of classical spins, in analogy with the homogeneous slow and fast flavor cases. In contrast, the SW system is not integrable (it does not possess Gaudin invariants) so that, while two-beam pendulum solutions are inevitable, they do not extend to a multi-angle system. We develop a taxonomy of all known nonlinear collective flavor solutions, explaining the overlap between categories and their differences.
}

\begin{document}
\maketitle
\flushbottom

\clearpage

\section{Introduction}

Neutrino-neutrino refraction~\cite{Pantaleone:1992eq} leads to collective flavor evolution~\cite{Samuel:1993uw, Samuel:1995ri}, with consequences for core-collapse supernovae and neutron-star mergers that remain incompletely understood (see Refs.~\cite{Duan:2005cp, Duan:2006an, Tamborra:2020cul, Volpe:2023met, Johns:2025mlm, Raffelt:2025wty} for reviews). Our own perspective has evolved to view the system as a neutrino plasma~\cite{Fiorillo:2024bzm, Fiorillo:2024uki, Fiorillo:2024pns, Fiorillo:2025ank, Fiorillo:2025zio,Fiorillo:2025gkw}---essentially a new state of matter---with many compelling analogies to the traditional electromagnetic plasma. In this framework, collective flavor motion is encapsulated in flavor waves (and their quanta flavomons~\cite{Fiorillo:2025npi, Fiorillo:2025kko}), independent degrees of freedom analogous to plasmons, which can redistribute flavor among neutrinos through neutrino-flavomon interactions. Quasi-linear theory for the backreaction of flavor waves on the system~\cite{Fiorillo:2024qbl, Fiorillo:2025npi} is perhaps the most promising framework for a quantitative and theoretical understanding of the evolution once instabilities of the flavor field begin to develop (see also Ref.~\cite{Johns:2025yxa} for a recent study along similar lines).

On the other hand, the original appeal of collective flavor evolution was motivated by a different emphasis (see Refs.~\cite{Duan:2009cd, Duan:2010bg} for early reviews): self-maintained coherence emerged as the most unexpected feature of dense neutrino gases---namely, highly regular motions in circumstances where rapid flavor equilibration would otherwise have been expected. Synchronized oscillations \cite{Samuel:1993uw, Kostelecky:1993yt, Kostelecky:1993ys, Kostelecky:1994dt, Pastor:2001iu}, nonlinear flavor waves \cite{Duan:2021woc}, flavor pendula \cite{Samuel:1995ri, Duan:2005cp, Hannestad:2006nj, Duan:2007mv, Raffelt:2011yb, Johns:2019izj,Padilla-Gay:2021haz, Fiorillo:2023mze, Fiorillo:2024dik}, potentially leading to spectral splits for adiabatic changes in the parameters \cite{Duan:2006an, Duan:2006jv, Fogli:2007bk, Raffelt:2007cb, Raffelt:2007xt, Dasgupta:2009mg}, and solitons~\cite{Fiorillo:2023hlk} are among the creatures inhabiting the traditional collective-flavor bestiary. Very recently, yet another species was added to this menagerie in the form of nonlinear single-wave (SW) solutions~\cite{Liu:2025muc}, which partly overlap with the previous cases, while also extending them in new directions.

Such highly regular exact solutions of the nonlinear equations are tantalizing, especially when only numerical methods or linearized approximations are otherwise available. It is, however, entirely unclear whether they play any practical role, since they typically rely on excessive symmetry assumptions; once these are relaxed, the regularity tends to dissipate into the broader phase space of possible excitations. Setting this fundamental doubt about the practical relevance of such very symmetric situations aside, the newly discovered SW solutions have nonetheless motivated us to develop a deeper understanding of their structure and how they fit into the broader picture of nonlinear exact solutions. 

\begin{figure}
    \centering
    \includegraphics[width=1.0\textwidth]{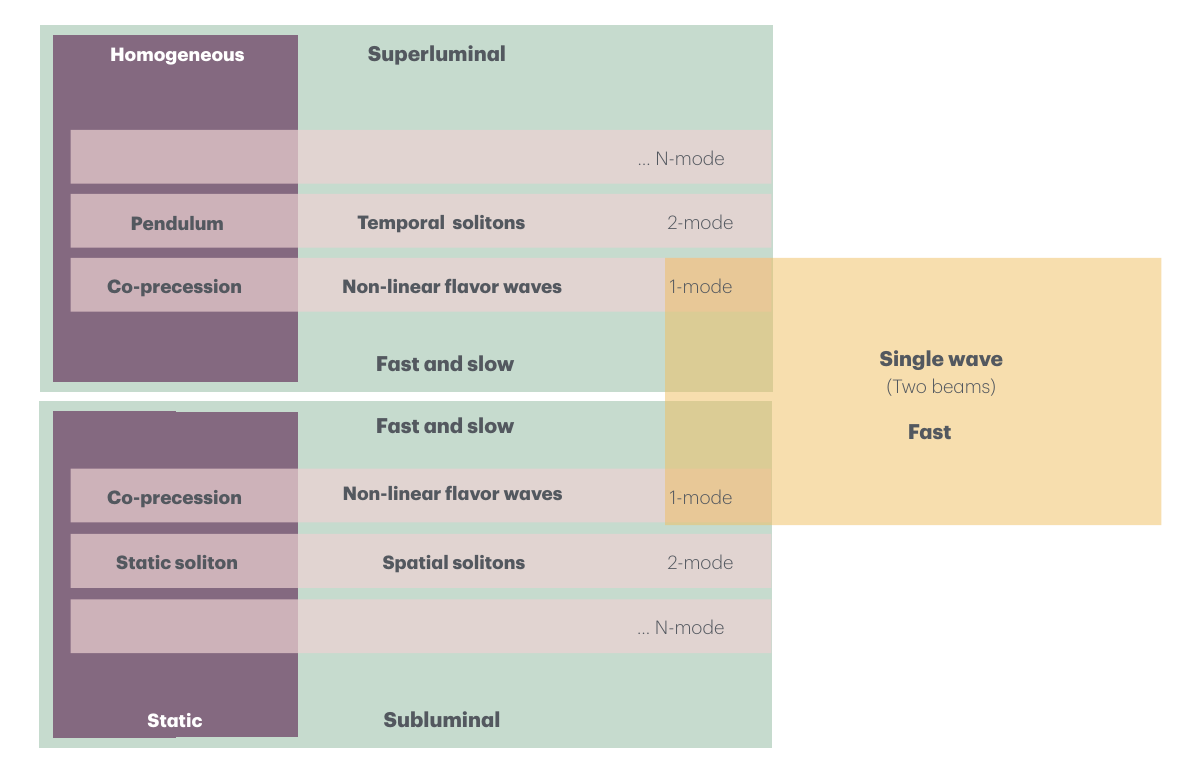}
    \caption{Classification of the known exact nonlinear solutions of collective flavor evolution. As usual, slow refers to flavor evolution driven by neutrino mass differences, whereas fast to the limit of vanishing masses. The inhomogeneous solutions (green and yellow boxes) are axisymmetric. The green block represents the single-coordinate solutions discussed in Sec.~\ref{sec:Single-Coordinate} which arise from homogeneous or static ones through a Lorentz boost.}\label{fig:scheme}
\end{figure}

The homogeneous slow and fast flavor systems are each equivalent to a collection of classical interacting spins. Such systems allow for $N$-mode coherent solutions, meaning that each spin evolves as a linear superposition of $N$ underlying modes \cite{Raffelt:2011yb}. The existence of such regularities is guaranteed by the integrability of these systems, which itself follows from their large number of so-called Gaudin invariants. Coprecession solutions are of this form, for which all spins lie in a single plane and precess around an external direction with a common frequency, but also the slow and fast flavor pendulum, in which neutrinos start from a nearly flavor-diagonal state and undergo pendular oscillations. 

Thus far, the known exact inhomogeneous solutions were usually derived from these homogeneous ones effectively through Lorentz boosts~\cite{Fiorillo:2023hlk}. Nonlinear flavor waves are the boosted versions of coprecession solutions, whereas flavor solitons are the boosted versions of flavor pendula. There exist specular versions of these cases with the roles of space and time interchanged, such as a static solution in the form of a ``spatial pendulum'' and the boosted version of a subluminal soliton. Henceforth such specular solutions are understood to exist without explicit mention.

Our systematic classification is summarized in Fig.~\ref{fig:scheme}, where as usual \textit{slow} refers to collective flavor evolution driven by neutrino mass differences~\cite{Samuel:1993uw, Samuel:1995ri, Fiorillo:2024pns, Fiorillo:2025ank}, whereas \textit{fast} to the limit of vanishing masses~\cite{Sawyer:2004ai, Sawyer:2008zs,Izaguirre:2016gsx,Fiorillo:2024bzm,Fiorillo:2024uki}. The diagram is most easily read beginning at its upper-left corner, corresponding to purely homogeneous systems, which possess some of the most widely known exact solutions: coprecession and flavor pendula. These solutions exist both for the slow, single-angle case and for the fast, axially symmetric case. Solutions with higher degrees of complexity can be constructed, although they are much less studied. A Lorentz boost produces superluminally-moving counterparts such as superluminal nonlinear flavor waves and superluminal solitons. 

One cannot, of course, predict if new exact solutions may be found in the future. What we can provide is a systematization of the criteria leading to exact solutions. Ultimately, these require a small number of degrees of freedom. This may happen either by construction, in few-beam systems, as in the SW case, or via the existence of conservation laws constraining motion to a limited phase space. At present, such conservation laws have only been found for mechanical system depending on a single coordinate: they correspond to the Gaudin invariants which characterize the flavor solitons.

The main focus of our present study is the new category of SW solutions~\cite{Liu:2025muc}---the yellow block in Fig.~\ref{fig:scheme}---in which the flavor-diagonal occupation numbers are perfectly homogeneous, while the off-diagonal coherence varies as a plane wave with a single wave number. This structure cannot arise from a Lorentz transformation of a purely homogeneous solution unless the occupation numbers are not only homogeneous, but also static. In this special case, for a moving observer, they still look homogeneous and static, whereas the field of flavor coherence looks homogeneous for a suitably moving observer, implying that these solutions coincide with the usual nonlinear flavor waves as indicated by the overlap of the yellow block with the usual nonlinear flavor waves in the green block.

Therefore, a truly new solution arises only when the occupation numbers are homogeneous but not static, and the flavor coherence  varies spatially like a plane wave, but the time variation is not harmonic. In particular, this includes situations when all polarization vectors begin aligned and develop an instability. Beyond the linear regime, this would be a flavor pendulum. While the authors of Ref.~\cite{Liu:2025muc} already noted that for the simplest two-beam case one finds a flavor pendulum, we conclude that such a solution inevitably exists for any two-beam system, but that it cannot be extended to multiple angles, in contrast to what happens in the fast system. The underlying reason is that the SW system of interacting classical spins is not integrable as it does not support Gaudin invariants.

The SW system therefore can be studied from different perspectives. It can be seen as the nonlinear counterpart of a single wave of the general kinematical fast flavor system. We will explore this perspective in Paper~II of this series~\cite{Fiorillo:2026vfo}, where we focus especially on the case of a weak instability. In this well-motivated case, even a continuous system still leads effectively to nonlinear saturation in the form of a flavor pendulum. In the present Paper~I, on the other hand, we focus on the SW system as a mechanical system of interacting classical spins, and compare its properties with those of the well-known homogeneous slow and fast systems. In particular, this concerns the inevitability of a two-beam flavor pendulum, its different formulations, the issue of Gaudin invariants, and the connected question if the two-beam flavor pendulum can be generalized to a continuous distribution of modes as the nonlinear incarnation of an instability. This is possible for the slow and fast systems, but not for the SW system, where the two-beam flavor pendulum stands by itself.

To set the stage, we begin in Sect.~\ref{sec:QKE} with the usual kinetic equations in different limits, and notably recall the single-angle slow system and the axisymmetric fast system. In Sect.~\ref{sec:spins}, we consider their homogeneous limits, leading to mechanical systems of interacting classical spins, and likewise derive analogous equations for the SW system. We discuss the integrability based on the existence or not of Gaudin invariants. In Sec.~\ref{sec:Single-Coordinate} we recall how $N$-mode coherent solutions of homogeneous systems can be turned to exact inhomogeneous solutions by a Lorentz boost. In Sect.~\ref{sec:coprecession} we consider coprecession solutions, which for the fast and slow cases become nonlinear flavor waves after a Lorentz boost, whereas for the SW system, they already are nonlinear flavor waves with the given wave number. In Sect.~\ref{sec:pendulum} we turn to the most interesting solutions arising from instabilities of an initial ensemble of flavor-aligned polarization vectors. In the slow and fast homogeneous systems, these connect, in the nonlinear regime, to pendulum-like solutions that can be analytically understood. In Sect.~\ref{sec:SW-unstable} we discuss unstable configurations of the SW system, which for two beams behave like a flavor pendulum, but cannot be extended to multiple angles. In other words, a SW instability for a continuous system does not possess a nonlinear pendulum solution due to its lack of Gaudin invariants, except approximately for the case of a weak instability to be explored in Paper~II \cite{Fiorillo:2026vfo}. Finally, Sect.~\ref{sec:discussion} is given over to a discussion and summary.

\section{Kinetic equations}

\label{sec:QKE}

To set the stage, we first report the usual mean-field kinetic equations for a two-flavor setup. We then reduce the equations to axial symmetry, and finally reduce them to two limiting single-parameter cases. One is the slow system, which arises as the single-angle limit, so that the neutrino modes depend only on energy. The other is the fast system, which arises from neglecting neutrino masses, thereby removes energy, and in axial symmetry leaves only the velocity projection on the symmetry axis as a single parameter.

\subsection{General case}

In the mean-field approach that we use throughout, the occupation number $f_{\bp}$ of each neutrino mode with momentum $\bp$ becomes a $2\times2$ matrix in flavor space, $\varrho_{\bp}$, where the diagonal entries are the occupation numbers for the two flavors, while the off-diagonal entries encode the degree of flavor coherence. The matrix $\varrho_{\bp}$ is Hermitean and thus can be expressed in terms of Pauli matrices $\sigma_i$ and a polarization vector $\bP_{\bp}$ in flavor space through
\begin{equation}
    \varrho_{\bp}
    =\frac{1}{2}\bigl({\rm Tr}\,\varrho_{\bp}+\bP_{\bp}\cdot\bsigma\bigr)
    =\frac{1}{2}\begin{pmatrix}
        f_{\bp}+P^z_{\bp} & P^x_{\bp}-iP^y_{\bp} \\
        P^x_{\bp}+iP^y_{\bp} & f_{\bp}-P^z_{\bp}
    \end{pmatrix}
    =\frac{f_\bp}{2}\,\mathbb{1}+\frac{1}{2}\begin{pmatrix}
        D_{\bp} & \Psi^*_{\bp} \\
        \Psi_{\bp} & -D_{\bp}
        \end{pmatrix},
\end{equation}
where $\bP$ is a vector in flavor space, whereas $\bp$ is one in coordinate space. For the off-diagonal term, that represents the field of flavor coherence, we will use the notation $\Psi_{\bp}=P^x_{\bp}+iP^y_{\bp}$. Furthermore, the on-diagonal element corresponds to the difference in occupation number of the two flavors, and hence we will refer to this difference in flavor lepton number (DLN) as $D_{\bp} =P^z_{\bp}$.

The usual mean-field kinetic equations~\cite{Dolgov:1980cq, Rudsky, Sigl:1993ctk, Fiorillo:2024fnl, Fiorillo:2024wej} for ultrarelativistic neutrinos---in the absence of matter and without symmetry assumptions---are
\begin{eqnarray}\label{eq:QKE-1}
    \left(\partial_t+\bv\cdot\bpartial_\br\right)\varrho_\bp
    &=&-i[\sH_\bp,\varrho_\bp]
    \nonumber\\[1ex]
    \text{with}\quad
    \sH_\bp&=&\frac{\sM^2}{2E}+\sqrt{2}\GF\int\frac{d^3\bp'}{(2\pi)^3}\,(1-\bv\cdot\bv')\bigl(\varrho_{\bp'}-\overline\varrho_{\bp'}\bigr),
\end{eqnarray}
where $\sM$ is the neutrino mass matrix, $E=|\bp|$ the energy,
$\bv=\bp/E$ the velocity, which in the ultrarelativistic limit is a unit vector, and $\overline\varrho_\bp$ is the density matrix for antineutrinos. They follow the same equation with a change of sign for $\sM^2/2E$. In terms of polarization vectors, the same equations are
\begin{eqnarray}\label{eq:general-precession}
    \left(\partial_t+\bv\cdot\bpartial_\br\right)\bP_\bp
    &=&\bH_\bp\times\bP_\bp
    \nonumber\\[1ex]
    \text{with}\quad
    \bH_\bp&=&\omega_E\bB+\sqrt{2}\GF\int\frac{d^3\bp'}{(2\pi)^3}(1-\bv\cdot\bv')\bigl(\bP_{\bp'}-\overline\bP_{\bp'}\bigr),
\end{eqnarray}
where $\omega_E\bB$ is the polarization vector for the matrix $\sM^2/2E$. Specifically, $\bB$ is traditionally chosen to be a unit vector in flavor space that defines the mass direction, whereas $\omega_E=\delta m^2/2E$ encodes the neutrino mass-squared difference. The form of Eq.~\eqref{eq:general-precession} shows that the motion for every $\bP_\bp$ is an instantaneous precession, keeping the length of $\bP_\bp$ fixed along the trajectory.

\subsection{Flavor isospin convention}

The structure of these equations suggests the often-used flavor isospin convention, where antineutrino modes are interpreted as modes with negative $E$ or rather negative $\omega_E$ and the polarization vectors for antineutrinos point in the opposite direction. In this case, the full phase space is covered by the energy $-\infty<E<+\infty$ and the direction of motion~$\bv$, which is a unit vector in the usual ultrarelativistic limit. Moreover, since energy only enters the equations through $\omega_E$, it is often convenient to use $\omega_E$ as a direct label for the modes so that the equations of motion (EoMs) become
\begin{eqnarray}\label{eq:general-precession-flavorisospin}
    \left(\partial_t+\bv\cdot\bpartial_\br\right)\bP_{\omega,\bv}
    &=&\bH_{\omega,\bv}\times\bP_{\omega,\bv}
    \nonumber\\[1ex]
    \text{with}\quad
    \bH_{\omega,\bv}&=&\omega\bB+\sqrt{2}\GF\int_{-\infty}^{+\infty }d\omega'\int\frac{d^2\bv'}{(2\pi)^3}(1-\bv\cdot\bv')\,\bP_{\omega',\bv'},
\end{eqnarray}
where the Jacobian for the $E\to\omega$ transformation has been absorbed in the length of $\bP_{\omega,\bv}$.

\subsection{Axial symmetry}

We always assume axial symmetry for both the initial setup and the solutions, implying that we no longer envision ``beams'' of neutrinos, but rather ``cones'' that are characterized by a common velocity projection $v=\cos\theta$ on the symmetry axis. Absorbing a suitable phase-space factor in the normalization of the polarization vectors, the EoMs become
\begin{eqnarray}\label{eq:precession-axial}
    \left(\partial_t+v\partial_r\right)\bP_{\omega,v}
    &=&\bH_{\omega,v}\times\bP_{\omega,v}
    \nonumber\\[1ex]
    \text{with}\quad
    \bH_{\omega,v}&=&\omega\bB+\sqrt{2}\GF\int_{-\infty}^{+\infty }d\omega'\int_{-1}^{+1}dv\, (1-vv')\,\bP_{\omega',v'}.
\end{eqnarray}
Somewhat surprisingly, the azimuthal integration takes the current-current factor $1-\bv\cdot\bv'$ to the structure $1-vv'$, implying that neutrinos with the same $v$ have nonvanishing refraction on each other. We recall that a given $-1<v<+1$ represents a cone, not a beam, and neutrinos within a given cone generally intersect each other at nonvanishing angles, leaving a net refraction effect among neutrinos within such a cone.

\subsection{Slow flavor system}

One can reduce this multi-angle and multi-energy system to a single-parameter system in different ways. One way is to reduce the angular distribution to a single $v$ so that all nontrivial effects derive from the distribution of vacuum oscillation frequencies and thus from neutrino masses, leading to $ (\partial_t+v\partial_r)\bP_\omega =(\omega\bB+\mu\bP_0)\times\bP_\omega$, where $\mu=\sqrt{2}\GF n_\nu$ is a measure of the neutrino-neutrino interaction strength. The total polarization vector is defined as $\bP_0=\int d\omega\,\bP_\omega$, which for now remains a function of both $r$ and $t$. All $\bP_\omega$ are understood to refer to the chosen single direction $v$.

\eject

A further simplification is achieved by making all energies as well as time and space dimensionless by absorbing $\mu$ in their definition, so that finally the slow system is
\begin{equation}\label{eq:precession-slow}
    \left(\partial_t+v\partial_r\right)\bP_w =(w\bB+\bP_0)\times\bP_w,
\end{equation}
where $w=\omega/\mu$. It will prove useful to write these equations also in component form
\begin{subequations}\label{eq:generic_slow_eom}
\begin{eqnarray}\label{eq:generic_slow_eom_1}
    \left(\partial_t+v\partial_r\right) D_w&=&\frac{i}{2}\left(\Psi_0\Psi_w^*-\Psi_0^*\Psi_w\right),
    \\
    \left(\partial_t+v\partial_r\right)\Psi_w&=&\,i\,\left[(w+
    D_0)\Psi_w-\Psi_0D_w\right],\label{eq:generic_slow_eom_2}
\end{eqnarray}    
\end{subequations}
where $D_0=\int dw\,D_w$ and $\Psi_0=\int dw\,\Psi_w$.

\subsection{Fast flavor system}

The purest form of collective flavor evolution is the fast case, defined by the limit of massless neutrinos so that the $\sM^2/2E$ term disappears. In this case, energy does not appear explicitly and can be integrated out, leading to another single-parameter system, where the modes depend only on the angle $v$, but not on energy. After absorbing phase-space factors and energy scales in the normalization of the polarization vectors and in the units of space and time, the EoMs become
\begin{equation}\label{eq:precession-fast}
    (\partial_t+v\partial_r) \bP_v=(\bP_0-v\bP_1)\times \bP_v,
\end{equation}
describing the precession of each polarization vector $\bP_v$ around the collective moments $\bP_0=\int dv\,\bP_v$ and $\bP_1=\int dv\,v\bP_v$. The general notation for the moments is $\bP_n=\int dv\,v^n\bP_v$, which also applies to the components $D_v$ and $\Psi_v$. In component form, the EoMs are
\begin{subequations}\label{eq:generic_fast_eom}
\begin{eqnarray}\label{eq:generic_fast_eom_1}
    (\partial_t+v\partial_r) D_v&=&\frac{i}{2}\left[(\Psi_0-v\Psi_1)\Psi_v^*-(\Psi_0^*-v\Psi_1^*)\Psi_v\right],
    \\
    (\partial_t+v\partial_r)\Psi_v&=&\,i\,\left[(D_0-vD_1)\Psi_v-(\Psi_0-v\Psi_1)D_v\right].\label{eq:generic_fast_eom_2}
\end{eqnarray}    
\end{subequations}
From the precession form of the equations it is evident that $|\bP_v|^2=D_v^2+|\Psi_v|^2$ is conserved along the trajectory of the neutrino mode.

\section{Systems of interacting spins}

\label{sec:spins}

The best known exact solutions pertain to situations of perfect homogeneity. The partial differential nature of the problem disappears after dropping the advection term in the transport operator on the left-hand side. The system is no longer kinetic, with particles streaming along different trajectories, but instead mechanical, with different polarization vectors (flavor~spins) coupled to one another by the refractive all-to-all interaction.  Intriguingly, keeping the advection term for a fixed wave number also reduces the system to such an ensemble of mechanical spins---the new SW system. We now consider these spin systems and discuss under which conditions they are integrable.

\subsection{Homogeneous slow and fast systems}

If the density matrix is assumed to be perfectly homogeneous, depending only on time, the EoMs for the slow system simplify to 
\begin{equation}\label{eq:EoM-homogeneous-slow}
    \dot{\bP}_w=(w\bB+\bP_0)\times \bP_w
\end{equation}
and for the fast system to
\begin{equation}\label{eq:EoM-homogeneous-fast}
    \dot{\bP}_v=(\bP_0-v\bP_1)\times \bP_v,
\end{equation}
where we now denote by a dot the derivative with respect to time. 

These can be seen as the EoMs for systems of interacting classical spins $\bP_w$ or $\bP_v$ with $\bP_0$ being the total angular momentum. These precession equations descend from the classical Hamiltonians
\begin{equation}\label{eq:Hamiltonians}
  H_{\rm slow}=\bB\cdot\bP_1+\frac{1}{2}\bP_0^2
  \quad\text{and}\quad
  H_{\rm fast}=\frac{1}{2}(\bP_0^2-\bP_1^2),
\end{equation}
using Poisson brackets for the classical angular momentum variables $\bP_w$ or $\bP_v$. Moreover, in the slow case, the projection of $\bP_0$ on $\bB$ is conserved, whereas in the fast case, even $\bP_0$ itself is conserved. The $z$-direction in flavor space is defined by $\bB$ in the slow case and by $\bP_0$ in the fast case.

\subsection{Single-wave fast system}

The physical nature of the recently discovered SW solutions \cite{Liu:2025muc} is most easily understood if one begins with the EoMs of the fast system in component form of Eq.~\eqref{eq:generic_fast_eom}. The starting point is to seek a solution of flavor coherence $\Psi_v(r,t)$  in the form of a single spatial wave $\Psi_v(t)\,e^{iK r}$, keeping a general temporal variation, for which we use the same symbol $\Psi_v$. Inserting this ansatz into Eq.~\eqref{eq:generic_fast_eom_1} reveals that the right-hand side no longer has a spatial dependence so that it is consistent to assume that $D_v$ is homogeneous. Therefore, the EoMs for such a spatial single wave are
\begin{subequations}\label{eq:SW_eom}
\begin{eqnarray}\label{eq:SW_eom_1}
    \dot D_v&=&\frac{i}{2}\left[(\Psi_0-v\Psi_1)\Psi_v^*-(\Psi_0^*-v\Psi_1^*)\Psi_v\right],
    \\
    \dot\Psi_v&=&\,i\,\left[(D_0-vD_1-vK)\Psi_v-(\Psi_0-v\Psi_1)D_v\right],\label{eq:SW_eom_2}
\end{eqnarray}    
\end{subequations}
where both $D_v$ and $\Psi_v$ are functions of time alone. While these equations descend from a spatial single wave, the variables themselves no longer depend on space---we have found once more a single-parameter mechanical system.

We can easily understand this result, since the factor $e^{i K r}$ appearing for the off-diagonal components of the density matrix corresponds to a spatial rotation of the flavor axes along the flavor $z$-direction with a rate $K$. Therefore, if the density matrix is defined in a spatially corotating flavor frame, the EoMs become
\begin{equation}\label{eq:SW-precession}
    \dot\bP_v=[\bP_0-v(\bP_1+\bK)]\times\bP_v,
\end{equation}
where $\bK$ is a vector in flavor space (not in coordinate space) defined to point in the flavor direction. The flavor direction is defined by the initial $\bP_0$, which however  is no longer conserved, so it is $\bK$, not $\bP_0$, that provides an external vector in flavor space, in analogy to $\bB$ in the slow system. These EoMs descend from the classical Hamiltonian
\begin{equation}\label{eq:Hamiltonian-SW}
  H_{\rm SW}=\frac{1}{2}(\bP_0^2-\bP_1^2)-\bK\cdot\bP_1,
\end{equation}
which also implies that there is a conserved energy. Therefore, the mathematical structure of the problem is quite analogous to the EoMs of the slow and fast homogeneous systems Eqs.~\eqref{eq:EoM-homogeneous-slow} and~\eqref{eq:EoM-homogeneous-fast} and their Hamiltonians Eq.~\eqref{eq:Hamiltonians}.

We stress, however, that this analogy is purely formal in the following sense.
The physical interpretation of the polarization vectors $\bP_v(t)$ of the SW system is mixed in that the flavor-diagonal part $D_v(t)$ is a homogeneous function, whereas $\Psi_v(t)$ is interpreted as the Fourier coefficient of the SW of flavor coherence. Therefore, a specific structure for the inhomogeneous solution is hard-wired into these equations. Formally, however, we have recovered a mechanical, not kinetic, system of interacting spins.

\subsection{Single-wave slow system}

The same ansatz can be made for the single-angle slow system described by the EoMs in component form of Eq.~\eqref{eq:generic_slow_eom}. By analogous arguments, one finds yet another mechanical system with the EoMs
\begin{equation}\label{eq:SW-precession-slow}
    \dot\bP_w=[w\bB+\bP_0-v\bK]\times\bP_w,
\end{equation}
where $v$ refers to the single occupied angle mode. However, by construction, both $\bB$ and $\bK$ point in the $z$-direction, which here is the mass direction, not the flavor direction. Therefore, we may instead write, assuming $\bB$ is a unit vector, 
\begin{equation}\label{eq:SW-precession-slow-2}
    \dot\bP_w=[(w-v K)\bB+\bP_0]\times\bP_w,
\end{equation}
implying a simple transformation of the spectrum of oscillation frequencies, i.e., a polarization vector labeled with $w$ actually has effective vacuum oscillation frequency $w-v K$. Otherwise the structure is the same as for any slow system and so we will not study it any further. With SW system we will always mean the fast case.

\subsection{Integrability and Gaudin invariants}

In spite of these dramatic simplifications, with the solutions now depending only on time and the system being therefore mechanical, generally we still cannot expect regularity. Turning first to the slow system, the range of $w$ forms a continuum that could cause decoherence by dephasing different modes. Instead, self-maintained coherence can appear, equivalent to a small number of underlying modes---the solution can live in a subspace of very small dimensionality. Every $\bP_w(t)$ can be a linear superposition of the constant vector $\bB$ and $N$ linearly independent functions $\bP_i(t)$ with $i=1,\ldots, N$, in which case the solution is called $N$-mode coherent \cite{Raffelt:2011yb}.

The reason for such coherence stems primarily from the fact that this system is completely integrable. The polarization vectors $\bP_w$ are Hamiltonian variables corresponding to two degrees of freedom, since their lengths $|\bP_w|$ are conserved. However, the EoMs also admit another set of conserved invariants for each mode that were first discovered by Gaudin \cite{Gaudin:1976sv}, whereas in the neutrino flavor context they were introduced in Ref.~\cite{Pehlivan:2011hp} and extensively used in Ref.~\cite{Fiorillo:2023mze}. The Gaudin invariants are most easily formulated in terms of the Lax vectors
\begin{equation}
    \bL_u=\bB+\sum_{w\neq u}\frac{\bP_w}{u-w},
\end{equation}
where here $w$ represents a set of discrete frequencies, whereas most generally, $u$ could be any complex number. If the frequencies $w$ form a continuum, the summation is replaced by an integral $\int dw$, but then needs a regularization of the pole at $u=w$, for example in the form of the Cauchy principal value, especially when $u$ is limited to real values.

It is straightforward to verify that the quantities $I_w=\bL_w\cdot \bP_w$ are constants of the motion, the celebrated Gaudin invariants. They are often also called Gaudin Hamiltonians because each $I_w$ can be seen as the Hamiltonian for a system of a ``central spin'' coupled to a set of ``environmental'' ones. The Gaudin invariants can be combined to form other invariants, such as the original Hamiltonian, and allow one to interpret the same system of interacting spins in many different ways.

For a Hamiltonian system, the existence of one invariant for each degree of freedom implies the complete integrability of the system. The most general solutions must be conditionally periodic, i.e., a superposition of periodic solutions in each of the modes, although the overall solution itself need not be periodic. However, under many conditions the fast and slow homogeneous systems can exhibit even more regular forms of motion, with exact periodicity, corresponding to the $N$-mode coherent solutions anticipated earlier.

While the original Gaudin invariants refer to a system of spins interacting with each other and with an external $\bB$ field, we have shown earlier that an analogous construction pertains to the fast system that lacks an external field, but instead owns the total angular momentum $\bP_0$ as a conserved vector that plays the same role \cite{Fiorillo:2023mze}. For the fast system, the Lax vectors are
\begin{equation}
    \bL_u=\sum_{v\neq u}\frac{v\bP_v}{u-v},
\end{equation}
once more leading to Gaudin invariants of the form $I_v=\bL_v\cdot \bP_v$.
The similarity between the slow and fast systems is not accidental: in fact, the slow and fast homogeneous systems can be mathematically mapped to one another~\cite{Fiorillo:2023mze}. In the fast system, an $N$-mode coherent solution would be one where each $\bP_v$ is a linear superposition of $N+1$ underlying modes, one of them simply being the conserved $\bP_0$.

The existence of Gaudin invariants also leads to the conservation of other fundamental quantities. For the slow system, $\bP_0$ has conserved length. For the fast system, we find that $\bP_0$ is a constant of motion, whereas $\bP_1$ has conserved length. If we write the EoMs in terms of Hamiltonian vectors in the form $\bP_w=\bH_w\times\bP_w$ or $\bP_v=\bH_v\times\bP_v$, in both cases these invariants imply that $\bH_w=w\bB+\bP_0$ in the slow case and $\bH_v=\bP_0-v\bP_1$ in the fast case are always linear combinations of vectors which maintain a fixed length during the whole evolution.

\subsection{Gaudin invariants of the single-wave system?}

One may be tempted to think that the SW system of interacting spins, with its apparent similarities to the slow and fast systems, should also support Gaudin invariants, but this is not the case. We search for them in the general form
\begin{equation}
    I_v=\bK\cdot\bP_v+\sum_{v'\neq v}\alpha_{v v'}\bP_{v'}\cdot \bP_v.
\end{equation}
Using the EoMs, the time derivative becomes
\begin{eqnarray}\label{eq:dIvdt}
    \dot{I}_v&=&\bK\cdot\biggl[\sum_{v'}(\bP_{v'}\times\bP_v)(1-vv')
    -\sum_{v'\neq v}\alpha_{v v'}(\bP_{v'}\times\bP_v)(v'-v)
\biggr]
     \nonumber\\
    &&{}+\bP_v\cdot\!\!\sum_{v',v''\neq v}\alpha_{v v'}(\bP_{v''}\times \bP_{v'})v'' (v-v').
\end{eqnarray}
For the terms proportional to $\bK$ to drop out, we need $\alpha_{vv'}=(1-v v')/(v'-v)$, but with this choice, we directly see that the second line in Eq.~\eqref{eq:dIvdt} does not disappear, and instead $\dot{I}_v=(\bP_0\times\bP_1)\cdot\bP_v$. Hence, the Gaudin invariants do not exist, spoiling the exact integrability of the model, at least in the Gaudin form. Of course, this does not preclude the existence of more complicated invariants, nonlinear in the products of the different polarization vectors, perhaps leading to integrability. However, the lack of Gaudin invariants is enough to explain why the SW system is intrinsically different from, and less regular than, its homogeneous counterparts.

Another hint of this conclusion comes from the observation that neither $\bP_0$ nor $\bP_1$, nor their lengths, are conserved. Except for the $\bP_v$ themselves, there is apparently no moment of the angular distribution whose evolution is a pure precession with constant length.

\section{Single-coordinate solutions}
\label{sec:Single-Coordinate} 

Homogeneous solutions (in either time or a one-dimensional spatial coordinate) were studied for many years as proxies for more complicated situations. However, such solutions are typically not stable against spontaneous symmetry breaking: a solution need not inherit the symmetries of the initial state \cite{Raffelt:2007yz, Raffelt:2013rqa}. In particular, beginning with what would be a homogeneous flavor pendulum, quickly fragments into spatial $\bk$ modes \cite{Mangano:2014zda}. A system with such extended degrees of freedom loses the symmetry protection of the Gaudin invariants and thus the underlying reason for the solution to be confined to a lower-dimensional space. 

As an extreme assumption, it was postulated that an unstable system would be ergodic~\cite{Johns:2023jjt}, filling all of phase space that is not explicitly forbidden by symmetries, although this idea leaves open of how \textit{a priori} one would exclude the existence of conserved quantities that may not be immediately apparent. In contrast to this idea, the neutrino plasma framework~\cite{Fiorillo:2024qbl,Fiorillo:2025npi} considers the evolution akin to a new form of plasma turbulence, with the $\bk$ modes evolving into a quasi-linear ensemble of flavor waves with randomized phases for weak instabilities, which interact nonlinearly as the instability strength grows. Ergodicity is in this sense disfavored by the observation that certain configurations which could become unstable according to the known conservation laws are actually stable~\cite{Fiorillo:2025kko}, with the instability being truly driven by a kinematical ability of the neutrinos to emit flavor waves.

Either way, it is intriguing that one can systematically find exact inhomogeneous solutions by boosting the earlier homogeneous (or static) $N$-mode solutions to another Lorentz frame. This idea can be practically implemented by seeking solutions that depend purely on a single combination of time and space, $\xi=r-Vt$, where the ``velocity'' $V$ is an arbitrary parameter~\cite{Fiorillo:2023hlk}. This general strategy is not new---it has long been followed in fluids and plasmas to identify stationary and exact solutions of nonlinear systems. In fluids, this approach naturally leads to the identification of \textit{solitons}, localized excitations which propagate with a fixed velocity $V$ thanks to the balance between dispersion and nonlinearity. In plasmas, the closest analogue are \textit{BGK modes}, which similarly correspond to spatial structures which are stationary in some moving frame, and which are supported by the trapping of electrons in the electric field of collective waves.

In the neutrino context, we showed that a very close analogue exists for the flavor field~\cite{Fiorillo:2023hlk}. Depending on the magnitude of the parameter $V$, we can identify two different physical classes of exact solutions. For $|V|>1$, one can always find a frame in which the solution depends only on $t'=t-r/V$. In this frame, the evolution depends only on time and is completely homogeneous: the well-known fast flavor pendulum belongs to this class. Instead, for $|V|<1$, one can always find a frame in which the solution is purely static $r'=r-Vt$. In our previous work~\cite{Fiorillo:2023hlk}, we focused only on a special class of single-coordinate solutions, which correspond to the fast flavor pendulum in the purely homogeneous case, and which we dubbed flavor solitons more generally, since they consist of a localized region of nonlinear flavor conversion steadily moving.

Pendulum-like solutions can have different properties, depending on initial conditions. If all polarization vectors are initially perfectly aligned, the system is in an unstable fixed point and requires a nonvanishing seed to grow exponentially. Mathematically, one can also consider a moving pendulum that began its motion in the infinite past and will return to the upright position in the infinite future. Such a non-periodic motion could be called an instanton, but we avoid this term due to its specific meaning in quantum field theory, and we opted instead for the terminology of a temporal soliton \cite{Fiorillo:2023hlk}. One can also construct exact periodic solutions, corresponding to a periodically swinging pendulum, where however the polarization vectors never perfectly align at any time.

Overall, the existence of $N$-mode coherent homogeneous solutions inevitably implies that from a moving frame, one observes a flavor field that varies in both space and time in a regular fashion. This conclusion applies both to slow and fast evolution. The Lorentz boost need not be along a symmetry axis of the original system so that in the moving frame, the neutrino angle distribution need not be axially symmetric. In practice, though, this conclusion may not be helpful in finding such solutions as typically one would ask for solutions in a given frame with a given angle distribution. If that distribution could be made axially symmetric or even isotropic by a boost is another question. If the distribution can be made symmetric enough by a boost, the $N$-mode coherent solutions exist. If regular space-time solutions can also exist without this symmetry property is unknown. 

Yet another question is the presence of ordinary matter that singles out a Lorentz frame. Earlier we have found that solitons are ``fragile'' in that they are destroyed by a matter current, i.e., if there is a relative velocity between the matter background and the soliton \cite{Fiorillo:2023hlk}. Therefore, in practice, the symmetry properties of the neutrino gas alone do not reveal the mathematical possibility of stable regular motion. However, here we do not further examine the impact of a matter background for regular solutions.

\section{Coprecession solutions and nonlinear flavor waves}

\label{sec:coprecession}

The simplest regular motion of all three mechanical systems is coprecession of all polarization vectors. For the slow and fast homogeneous system, these can be boosted to become nonlinear flavor waves, whereas for the SW system, a coprecession solution already is a nonlinear flavor wave, i.e., a solution that varies harmonically in both space and time. We here review some general properties of coprecession solutions.

\subsection{General properties}

All three mechanical spin systems allow for coprecession solutions, meaning that all $\bP_w$ or $\bP_v$ lie in a single plane that rotates around the $z$-direction with a certain frequency $\Omega_{\rm c}$. Within that corotating plane, all $\bP_w$ or $\bP_v$ are static relative to each other and therefore proportional to their coprecessing Hamiltonian vectors $\bH_w$ or $\bH_v$. Assuming the length distribution $P_w$ or $P_v$ is given, one can find self-consistency conditions that allow one to find $\Omega_{\rm c}$ and the required angles $\theta_w$ or $\theta_v$ relative to the $z$-direction, solutions that need not be unique. 

The physical interest in such configurations came about in the context of explaining spectral splits in the slow system, when the effective interaction strength $\mu$ began at some large value, where the $\bP_w$ were nearly aligned, to some small value, where they were aligned again, but the $z$-orientation of certain ranges had been swapped \cite{Duan:2006jv, Fogli:2007bk, Raffelt:2007cb, Raffelt:2007xt, Dasgupta:2009mg}. For the self-consistency conditions for a fixed $\mu$, we specifically refer to Ref.~\cite{Raffelt:2011yb}. In other words, all coprecession solutions of the slow system are characterized by their spectrum and common orientation at $\mu\to\infty$, to which they are connected by an adiabatic transformation. Indeed, $\mu\to\infty$, when they are all collinear, is a special case of coprecession in the form of synchronized oscillations, which is physically understood as all spins sticking to each other due to their strong mutual interactions, and can only precess together as one collective object around the external $\bB$ field~\cite{Pastor:2001iu}. In the fast and SW systems, there is no direct equivalent to this limit because $\mu$ is absorbed in the units of space and time, whereas $v$ is always in the range $[-1,+1]$.

\subsection{Fast system}

\label{sec:Coprecession-fast}

Since coprecession in the slow case has been abundantly discussed in the literature, we here pay explicit attention only to the fast system. The individual Hamiltonian vectors are $\bH_v=\bP_0-v\bP_1$. Since $\bP_0$ defines the $z$-direction, in the coprecessing frame one finds $H_v^z=D_0-v D_1-\Omega_{\rm c}$, where $D_0=P_0^z$ is positive by definition. We introduce the shifted coprecession frequency $\omega=D_0-\Omega_{\rm c}$, not to be confused with slow-system oscillation frequencies, to connect to previous studies of the fast system. Seeking its normal modes beginning with the EoMs in component form of Eq.~\eqref{eq:generic_fast_eom}, one assumes $\Psi_v(t)=\psi_v\,e^{-i\Omega t}$, where $\Omega=-\Omega_{\rm c}$, i.e., with our convention of $\Psi_v=P_v^x+iP_v^y$, a positive precession frequency corresponds to a negative normal-mode frequency. In the comoving frame, the amplitude $\psi_v$ is real, so that the transverse component is $H_v^\perp=-v\psi_1$, and the modulus is $|\bH_v|=\sqrt{(\omega-v D_1)^2+ (v\psi_1)^2}$. The components of the Hamiltonian unit vectors are finally 
\begin{equation}\label{eq:H-components-fast}
  \hat H_v^z=\frac{\omega-v D_1}{\sqrt{(\omega-v D_1)^2+ (v\psi_1)^2}}
  \quad\text{and}\quad
  \hat H_v^\perp=\frac{-v \psi_1}{\sqrt{(\omega-v D_1)^2+ (v\psi_1)^2}}.
\end{equation}
In the coprecession configuration, each $\bP_v$ is collinear with its $\bH_v$, but can point in the same or opposite direction, so that $\bP_v=g_v \hat\bH_v$, where $g_v=\pm P_v$ with $P_v=|\bP_v|$. In other words, the angle of the ray of $\bP_v$ relative to the $z$-direction is
\begin{equation}
  \theta_v=\arctan\,\frac{-v\psi_1}{\omega-v D_1}.
\end{equation}
Depending on the parameters of the solution, there can be a resonance on the interval $-1<v<+1$, simply meaning that $\bH_v$ is exactly transverse to the flavor axis and $\bP_v$ must point in the transverse direction.

Self-consistency requires that $\psi_0=\int dv\, g_v \hat{H}_v^\perp$ and $\psi_1=\int dv\, g_v v\,\hat{H}_v^\perp$. Moreover, $\psi_0=0$ because $\bP_0$ defines the $z$-direction, so that the two conditions are
\begin{subequations}
  \begin{eqnarray}\label{eq:Selfconsistency-fast-a}
   0&=&\int dv\,g_v\,\frac{v}{\sqrt{(\omega-v D_1)^2+ (v\psi_1)^2}},\\
   \label{eq:Selfconsistency-fast-b}
   -1&=&\int dv\,g_v\,\frac{v^2}{\sqrt{(\omega-v D_1)^2+ (v\psi_1)^2}}.
  \end{eqnarray}
The condition $D_1=\int dv\, g_v v\,\hat{H}_v^z$ is identically true if the two previous relations are fulfilled. In the condition $D_0=\int dv\, g_v \hat{H}_v^z=\int dv\, g_v (\omega-v D_1)/\sqrt{(\omega-v D_1)^2+ (v\psi_1)^2}$, the second term vanishes per Eq.~\eqref{eq:Selfconsistency-fast-a}, leaving us with a third condition
\begin{equation} \label{eq:Selfconsistency-fast-c}
  D_0=\omega \int dv\,\frac{g_v}{\sqrt{(\omega-v D_1)^2+ (v\psi_1)^2}}.
\end{equation}
\end{subequations}
From Eq.~\eqref{eq:Selfconsistency-fast-b} we see immediately that $g_v$ cannot be positive everywhere for a solution to exist, but on the other hand, $g_v=-P_v$ is not excluded. Anyway, the spectrum has no direct physical meaning because its sign is defined relative to $\hat\bH_v$ and thus, for a given $P_v$, is part of the solution. In other words, for the same $P_v$, different assumed sign functions $\epsilon_v$ with $\epsilon_v=\pm1$ such that $g_v=\epsilon_v P_v$ may lead to different solutions.

To solve the equations in practice for a given spectrum $g_v$, one can first introduce $\bar D_1=D_1/\omega$ and $\bar\psi_1=\psi_1/\omega$, so that $\omega$ drops out from Eq.~\eqref{eq:Selfconsistency-fast-a}, and then find a ``dispersion relation'' between $\bar D_1$ and $\bar \psi_1$. For a given pair $\{\bar D_1,\bar \psi_1\}$ we can then use Eq.~\eqref{eq:Selfconsistency-fast-b} to find 
\begin{equation}\label{eq:Selfconsistency-fast-omega}
   |\omega|=-\int dv\,g_v\,\frac{v^2}{\sqrt{(1-v \bar D_1)^2+ (v\bar\psi_1)^2}}.
  \end{equation}
Until this point, we have found $\{\bar D_1,\bar\psi_1,|\omega|\}$, if a solution exists at all, and finally use Eq.~\eqref{eq:Selfconsistency-fast-c} to find $D_0$, which by definition is positive, so that this equation also fixes the sign of $\omega$, so that finally we have found $\{D_0, D_1, \psi_1,\omega\}$ and thus all elements of the solution.

\subsection{Nonlinear flavor waves}

Coprecession solutions have the property that the projection of the polarization vectors on the flavor direction, $D_v$, is static, whereas the transverse components, in complex form the field of flavor coherence, $\Psi_v(t)$, evolves harmonically in the form $\Psi_v(t)=\psi e^{-i\Omega t}$, where $\Omega$ is a real frequency. Viewed by a moving observer, the diagonal piece $D_v$ remains static and homogeneous, whereas $\Psi_v$ must show a variation that is harmonic both in space and time and thus varies in the form $e^{-i(\Omega t- K r)}$, the definition of a nonlinear flavor wave.

Of course, a moving observer also sees a modified angle distribution. If the motion is along the symmetry axis, the new angle distribution is once more axisymmetric. Nonlinear flavor waves of this type, of course, can also be found directly for a given angle distribution, a feat first achieved by Duan, Martin, and Omanakuttan~\cite{Duan:2021woc}.

It is noteworthy that the EoM for the flavor field in component form Eq.~\eqref{eq:generic_fast_eom_2} is linear, meaning that for any solution $\{D_v,\Psi_v\}$ with static $D_v$, another solution is given by the same $D_v$ and any multiple of $\Psi_v$, notably if it is infinitesimally small. Therefore, for a given spectrum $D_v$, one can determine the dispersion relation to find consistent real-valued pairs $\{\Omega,K\}$, which then also applies to the corresponding nonlinear flavor waves, where the flavor field need not be small compared with $D_v$. In this sense, it is justified to speak of a dispersion relation for nonlinear flavor waves~\cite{Duan:2021woc}.

\subsection{Coprecession solutions of the single-wave system}

The structure of the spin-precession EoM of 
Eq.~\eqref{eq:SW-precession}, which again is of the form $\dot\bP_v=\bH_v\times\bP_v$ suggests that there can be coprecession solutions, meaning that all $\bP_v$ are static relative to each other so that, in particular, the length of both $\bP_0$ and $\bP_1$ are conserved. To construct explicit solutions directly from the single-wave EoMs, one can proceed as in Sect.~\ref{sec:Coprecession-fast}. The components of the Hamiltonian vectors are here $H_v^z=\Omega+D_0-v(K+D_1)$ and $H_z^\perp=\psi_0-v\psi_1$, with phases chosen such that $\psi_0$ and $\psi_1$ are real. With the usual shifted frequency and momentum, $\omega=\Omega+D_0$ and $k=K+D_1$, the unit vectors are
\begin{equation}
  \hat H_v^z=\frac{\omega-vk}{\sqrt{(\omega-v k)^2+ (\psi_0-v\psi_1)^2}}
   \quad\text{and}\quad
  \hat H_v^\perp=\frac{\psi_0-v \psi_1}{\sqrt{(\omega-v k)^2+ (\psi_0-v\psi_1)^2}}
\end{equation}
and an angle distribution relative to the $z$-axis of 
\begin{equation}
  \theta_v=\arctan\,\frac{\psi_0-v\psi_1}{\omega-v k}.
\end{equation}
Once more, there can be a resonance on the interval $-1<v<+1$, meaning that $\bH_v$ is exactly transverse to the flavor axis and $\bP_v$ must point in the transverse direction. For a given spectrum, solving the self-consistency conditions $\psi_0=\int dv\,g_v H_v^\perp$, $\psi_1=\int dv\,v\,g_v H_v^\perp$, $D_0=\int dv\,g_v H_v^z$, and $D_1=\int dv\,v\,g_v H_v^z$ is now much more complicated because these conditions cannot be solved one after the other as in Sect.~\ref{sec:Coprecession-fast}.

Another approach is to specify $D_v$, not $g_v$, and seek a consistent $\psi_v$. If $D_v$ is specified, we also know $D_0$ and $D_1$, and observe that Eq.~\eqref{eq:SW_eom_2} is linear in the flavor field. In other words, for every consistent solution $\psi_v$, any multiple is another solution. In particular, this applies to $\psi_v$ being infinitesimally small, so that $\omega$ and $k$ follow from the usual dispersion relation of the linearized fast system. As mentioned earlier, for given $D_v$, the linearized dispersion relation applies even when $\psi_v$ is not small compared to $D_v$.

In conclusion, the task of finding a nonlinear flavor wave is the same as finding a coprecession solution of the SW system. Explicit solutions of this type were first constructed in Ref.~\cite{Duan:2021woc} so that there is no need for further examples. Of course, not for every spectrum and every $K$ there is such a solution because for a chosen $K$, the eigenfrequency $\Omega$ can have an imaginary part, indicating an instability instead of a stable flavor wave.

Earlier we have found that the SW system does not support Gaudin invariants so that regular solutions are not guaranteed. On the other hand, coprecession is the most regular conceivable behavior. There is no contradiction because the absence of Gaudin invariants does not prove that there could not be special regular solutions after all. The special properties of the coprecession solution, being effectively a nonlinear flavor wave, which in turn is a boosted version of the homogeneous fast flavor coprecession, probably explains the existence of such one-mode coherent solutions of the SW system.

\section{Flavor pendulum---traditional fast and slow cases}

\label{sec:pendulum}

A more complicated case of regular motion is bimodal coherence, discovered and analytically solved in early studies of Samuel~\cite{Samuel:1995ri}, corresponding in our jargon to two-mode coherent solutions. The motion of any $\bP_w(t)$ or $\bP_v(t)$ is a linear combination of a constant vector and two dynamical spins, which together obey pendulum equations in a sense to be explained. For a homogeneous unstable slow or fast system, the unstable solutions of the linear system are continued to the nonlinear regime by the flavor pendulum~solutions.

\subsection{General picture}

After coprecession, the next more complicated form of regular motion consists of the flavor pendulum. The original observation was that the dynamics of two classical angular momenta $\bP_{w_1}$ and $\bP_{w_2}$ that follow the slow flavor EoMs is equivalent to a gyroscopic pendulum with a certain mapping of physical parameters~\cite{Hannestad:2006nj}. The same applies to the dynamics of three $P_{v_i}$ ($i=1,2,3$) obeying the fast flavor EoMs~\cite{Johns:2019izj, Padilla-Gay:2021haz, Fiorillo:2023mze, Fiorillo:2023hlk}. We will see that this conclusion also pertains to the single-wave system, where two $\bP_{v_1}$ and $\bP_{v_2}$ are once more enough because, like in the slow system, there is an external flavor vector called~$\bK$. One crucial  property of such systems is that the polarization vectors can develop large transverse components even if all of them begin almost perfectly aligned, i.e., the existence of instabilities.

One key property of the slow and fast systems is that the pendular dynamics is not limited to the case of a few discrete beams, but can arise for an entire continuum of $\bP_w$ or~$\bP_v$. Taking the fast system as an example, the solution for every $\bP_v(t)$ is a linear superposition of three spin vectors $\bL_{v_i}(t)$  with $i=1,2,3$, which themselves form a flavor pendulum consisting of three discrete modes. When the dynamics of the continuum $\bP_v(t)$ is equivalent to three discrete modes, we call the system 2-mode coherent as we do not count the preserved vector $\bP_0$. For the minimal slow pendulum, there is a conserved external vector $\bB$ and two dynamical spins. We will see that for the SW system, an extension to the continuous case is not possible, due to its presumed lack of integrability suggested by the absence of Gaudin invariants.

\subsection{Mapping to pendulum parameters}

What do we mean with pendulum dynamics? Considering a three-mode fast system, we mean that the three modes $P_{v_i}$ with $i=1,2,3$ can be linearly combined to form three other vectors $\bG$, $\bR$ and $\bJ$ that obey the pendulum EoMs
\begin{equation}\label{eq:pendulum}
  \dot\bG=0,
  \qquad
  \dot\bR=\bJ\times\bR,
  \qquad
  \dot\bJ=\bG\times\bR.
\end{equation}
Compared with a mechanical gyroscopic pendulum (or spinning top), $\bG$ is the gravitational field that exerts a force on the center of mass, $\bR$ is the vector connecting the point of support with the center of mass, and $\bJ$ is the total angular momentum. Conserved quantities are $\bR^2$ and thus the length of $\bR$, $\bJ\cdot\bG$ and thus $J_z$, the angular momentum along the direction of gravity, energy $\bG\cdot\bR+\bJ^2/2$, and $\bJ\cdot\bR$ and thus the spin, i.e., the projection of $\bJ$ on the radius vector, $\bS=\bJ\cdot\hat\bR$. The physical meaning of the second equation is that an infinitesimal motion of $\bR$ is a precession around the vector of total angular momentum, i.e., $\bJ$ generates infinitesimal rotations. The third equation means that gravity exerts a torque that tries to pull the center of mass down and leads to a change of the total angular momentum. All physical units such as the moment of inertia, the physical length of $\bR$, or the strength of the gravitational force have been absorbed in the lengths of the vectors as well as the unit of time. The natural pendulum frequency is $\lambda=\sqrt{G R}$ so that there remains an arbitrary factor between the length of $\bG$ and $\bR$, meaning that $\bG$ could be chosen as a unit vector, and then $R=|\bR|=\lambda^2$. The physical meaning of $\lambda$ is the oscillation frequency of a plane pendulum (no spin) in the stable configuration, executing a small oscillation.

For the fast system, one can identify $\bG=v_1v_2v_3\bP_0$, $\bR=\bP_1$, and $\bJ=\bP_2-(v_1+v_2+v_3)\bP_1$, where $\bP_n=\sum_i v_i^n \bP_{v_i}$ \cite{Padilla-Gay:2021haz}. For the slow system, the usual mapping is $\bG=\frac{1}{2}(w_1-w_2)\bB$, $\bR=\bP_{w_1}-\bP_{w_2}-\bG$, and $\bJ=\bP_0$. Replacing these interpretations of $\bG$, $\bR$, and $\bJ$ into Eq.~\eqref{eq:pendulum} verifies that one recovers the original EoMs for the fast or slow systems. We see that for the fast system, $\bR=\bP_1$ is simple and it is $\bP_1$ that moves like the center of mass of a mechanical pendulum. This remains true for the continuous system, whereas $\bJ$ is then an abstract quantity. For the slow system, it is $\bJ=\bP_0$ that is simple, whereas the radius vector, involving the external vector $\bB$, is less intuitive. Even for the continuous system, $\bJ=\bP_0$, i.e., the total angular momentum of the original system is also the total angular momentum of the pendulum, whereas the role of $\bR$ is even more abstract.

\subsection{Soliton solution}

The pendulum equations \eqref{eq:pendulum} can be solved analytically as shown in classical mechanics textbooks and in the neutrino context in Refs.~\cite{Raffelt:2011yb, Padilla-Gay:2021haz, Fiorillo:2023mze, Fiorillo:2023hlk}. In that context, we always begin with polarization vectors collinear with the $z$-direction, implying that this also holds for $\bR$ and $\bJ$. If there is an instability, some of the polarization vectors may not be perfectly aligned, but we always assume that there is no initial velocity for any of them. Therefore, the spin $\bS$ is initially identical with $\bJ$. We describe the pendular motion in terms of the polar coordinates of $\bR$, namely its angle $\vartheta$ against the $z$-direction and an azimuthal angle $\varphi$. These obey the equations
\begin{equation}\label{eq:pendulum-solution}
  \dot\varphi=\frac{2\lambda\sigma}{1+c_\vartheta}
  \quad\text{and}\quad
  \dot c_\vartheta^2=2\lambda^2(1-c_\vartheta)^2(1+c_\vartheta-2\sigma^2),
\end{equation}
where $c_\vartheta=\cos\vartheta$. The parameter $\sigma=|\bS|/2\lambda$ is the dimensionless pendulum spin, a parameter here defined to be positive. If the pendulum spins initially clockwise or counter clockwise makes no difference to the overall solution. For $\sigma=0$, there is no azimuthal motion and the $\vartheta$ motion is that of a plane pendulum. On the other extreme, for $\sigma\ge1$, the right-hand side of the second equation is negative for any $c_\vartheta$ so that there no solution other than $c_\vartheta=1$ and $\dot c_\vartheta=0$, the sleeping-top configuration, an upright position with the gyroscope spinning so fast it cannot fall.

In the unstable configuration ($0<\sigma<1$), and if initially $c_\vartheta$ is very close to one, the initial exponential growth phase for $\vartheta$ lasts arbitrarily long. The limiting solution would be an ``instanton,'' which we have preferred to call temporal soliton, where the motion starts in the infinite past and the pendulum returns to the upright position in the infinite future. Choosing $t=0$ as the instant of maximum excursion, it is explicitly \cite{Fiorillo:2023hlk}
\begin{equation}\label{eq:instanton}
  c_\vartheta=1-2(1-\sigma^2)\,{\rm sech}^2\left(\sqrt{1-\sigma^2}\lambda t\right),
\end{equation}
where ${\rm sech}(x)=2/(e^{-x}+e^{x})$. The pendulum solution is nearly the same, with a small excursion angle at the beginning, leading to the often-shown periodic pendulation. In other words, a temporal soliton is a pendulum with the waiting period between pendular swings being infinite instead of very long.

\subsection{Continuous system}

The main interest in the pendulum solution is its role in continuous slow or fast systems. Beginning with all polarization vectors nearly aligned with the $z$-axis, the distributions $P_v$ or $P_\omega$ need a crossing, which however alone does not guarantee an instability in the homogeneous cases, but now we assume one exists. In the linear regime, one can solve the dispersion relation and find the complex eigenfrequency $\Omega=\Omega_R+i\gamma$, where the positive imaginary part $\gamma$ is the growth rate. Considering the fast system, for the real part of the eigenfrequency, what actually matters is the shifted value $\omega_R=\Omega_R+D_0$. These parameters are connected with those of the underlying pendulum by the mapping \cite{Padilla-Gay:2021haz, Fiorillo:2023hlk}
\begin{equation}\label{eq:mapping}
  \sigma=\sqrt{\frac{\omega_R^2}{\omega_R^2+\gamma^2}}
  \quad\text{and}\quad
  \lambda=\sqrt{\omega_R^2+\gamma^2}.
\end{equation}
Expressed in terms of these parameters, the pendulum angle evolves as
\begin{equation}
  c_\vartheta=1-\frac{2\gamma^2}{\omega_R^2+\gamma^2}\,{\rm sech}^2\left(\gamma t\right),
\end{equation}
which is the angle of $\bP_1$ relative to the flavor axis.

One can then find the full solution for every $\bP_v(t)$, but we here only show the relatively simple expression for the $z$-component that we have worked out earlier~\cite{Fiorillo:2023hlk},
\begin{equation}
    D_v=g_v\left[1-\frac{v^2 D_1^2}{(\omega_R-vD_1)^2+\gamma^2}\,(1-c_\vartheta)\right],
\end{equation}
where $g_v$ is the spectrum which here is identical with the initial $z$-components, $g_v=P_v^z(0)$, that is, $|g_v|=P_v$ and by modulus identical with the length spectrum. We finally find explicitly
\begin{equation}\label{eq:Lorentzian_general}
    D_v=g_v\left[1-\frac{v^2D_1^2}{(\omega_R-vD_1)^2+\gamma^2}\,\frac{2\gamma^2}{\omega_R^2+\gamma^2}\,{\rm sech}^2\left(\gamma t\right)
    \right].
\end{equation}
This function has essentially a resonance structure, where $v_R=\omega_R/D_1$ can be seen as a resonance velocity, which  however need not lie in the range $-1\leq v\leq+1$.

Besides providing a complete analytical solution for the lepton number difference in the nonlinear regime of the pendulum, the structure of this expression gives us a new fundamental insight: the flavor pendulum has a characteristic resonant structure, affecting mostly neutrinos moving with the velocity $v=\omega_R/D_1$. This structure is a direct inheritance of the fundamental property of linear instabilities~\cite{Fiorillo:2024bzm, Fiorillo:2024uki, Fiorillo:2025npi}, that are caused by neutrinos resonantly moving in phase with the wave. Indeed, for a homogeneous wave, the resonance condition predicted by the theory of fast flavor instability is precisely $v_R=\omega_R/D_1$. For a weak instability, with $\gamma\ll D_1$, Eq.~\eqref{eq:Lorentzian_general} is, close to the resonance $v\sim \omega_R/D_1$, simply
\begin{equation}\label{eq:Lorentzian_weak}
    D_v=g_v\left[1-\frac{2\gamma^2}{(\omega-vD_1)^2+\gamma^2}\mathrm{sech}^2\gamma t\right].
\end{equation}
In Paper~II, we will see that the survival of the pendulum in the nonlinear regime is not exclusive of the homogeneous system, at least for weak instabilities. We will show that Eq.~\eqref{eq:Lorentzian_weak} remains valid for generic inhomogeneous, monochromatic instabilities, provided that the instability itself is weak. In this more generic case, the flavor pendulum is not exact, differently from the homogeneous case, where instead it exists for arbitrary strength of the instability, being protected by the existence of Gaudin invariants.

\section{Unstable solutions of the single-wave system}

\label{sec:SW-unstable}

The SW system has instabilities that can be found by the usual fast flavor dispersion relation, assuming the spectrum $D_v$ of initially aligned flavor spins is given. If the dispersion relation reveals unstable solutions for a given $K$, then a SW system for that $K$ is unstable. However, usually there is no continuation to the nonlinear regime in a similar form as discussed for the fast and slow systems. The only known nonlinear continuation arises in a two-beam system of arbitrary $v_1$ and $v_2$, which forms a two-mode flavor pendulum.

\subsection{Pendulum solution}

Looking at the SW system as a purely mechanical one, and following the logic of the fast flavor pendulum, it is clear that one needs a minimum of two dynamical modes $\bP_{v_1}$ and $\bP_{v_2}$ on top of the external vector $\bK$. In other words, we consider a two-beam system in analogy to Ref.~\cite{Liu:2025muc}, who however only chose the special case $v_{1,2}=\pm 1$. We thus ask if we can find linear combinations of the three vectors $\bK$, $\bP_{v_1}$, and $\bP_{v_2}$ to form a $\bG$, $\bR$ and $\bJ$ that obey the pendulum equations~\eqref{eq:pendulum}. We immediately identify $\bK$ to play the role of $\bG$, whereas $\bR=a_0\bK+a_1\bP_{v_1}+a_2\bP_{v_2}$ and $\bJ=b_0\bK+b_1\bP_{v_1}+b_2\bP_{v_2}$ are some as yet unknown linear combinations. Inserting these general expressions into the pendulum equations~\eqref{eq:pendulum}, one can match the coefficients and finds
\begin{equation}\label{eq:first_formulation}
  \bG=\bK,\quad
  \bR=v_1v_2\bK-(1-v_1v_2)\bP_1,\quad
  \bJ=-(v_1+v_2)\bK+(1-v_1v_2)\bP_0,
\end{equation}
where $\bP_0=\bP_{v_1}+\bP_{v_2}$ and $\bP_1=v_1 \bP_{v_1}+v_2 \bP_{v_2}$. The pendulum parameters are unique up to a rescaling of $R$ and $G$ such that $\lambda=\sqrt{GR}$ is fixed. Once we have found these expressions, they can be verified by replacing in the pendulum equations.

One can identify a pendulum also in a different manner. In the general ansatz $\bR=a_0\bK+a_1\bP_{v_1}+a_2\bP_{v_2}$, our previous solution is unique (up to a multiplicative factor) if we insist to involve $\bK$. On the other hand, if we set $a_0=0$, there are two more solutions $\bR=\bP_{v_1}$ or $\bP_{v_2}$ for which $|\bf R|$ is conserved. Focusing on one of them, its instantaneous precession is engendered by $\dot\bP_{v_1}=\bH_{v_1}\times\bP_{v_1}$ with its Hamiltonian vector $\bH_{v_1}=\bP_0-v_1(\bP_1+\bK)$. One can easily check that $\bH_{v_1}$ does not respect the correct equation for $\bJ$. However, in a frame rotating around $\bK$ with a frequency $v_2 |\bK|$, the EoMs of $\bP_{v_1}$ become $\dot{\bP}_{v_1}=\left[\bP_0(1-v_1 v_2)+(v_2-v_1)\bK\right]\times \bP_{v_1}=\bJ_{v_1}\times \bP_{v_1}$. In this corotating frame, the EoM for $\bJ$ is indeed $\dot{\bJ}_{v_1}=(1-v_1 v_2)(v_2-v_1)\bK\times \bJ_{v_1}$, so we find explicitly
\begin{equation}\label{eq:second_formulation}
  \bG_{v_1}=(1-v_1v_2)(v_2-v_1)\bK,
  \quad
  \bR_{v_1}=\bP_{v_1},
  \quad
  \bJ_{v_1}=(v_2-v_1)\bK+(1-v_1v_2)\bP_0.
\end{equation}
For the second beam, analogous expressions obtain by $1\leftrightarrow2$. Any of these formulations of the flavor pendulum describes the same collective motion of $\bP_{v_1}$ and $\bP_{v_2}$ and, in particular, the motion of the $z$-components is independent of the chosen frame and thus identical. 

Finally, we can express the pendulum parameters, as reviewed in Appendix~B of Ref.~\cite{Fiorillo:2023hlk}, for an initial configuration of polarization vectors that are collinear with the flavor direction, which is here defined by $\bK$. In our first formulation in Eq.~\eqref{eq:first_formulation}, the natural frequency is given by the parameter $\lambda^2=G^z R^z\vert_{t=0}=K[v_1 v_2 K-(1-v_1 v_2)D_1]$. In this formulation, $\lambda^2$ may also be negative: our expressions apply at the formal level in the same way. The spin is the initial $S=\bJ\cdot \bR/R=(1-v_1 v_2)D_0-(v_1+v_2)K$. Using the results of Ref.~\cite{Fiorillo:2023hlk}, we can relate these two parameters to the properties of the instability. The growth rate is $\gamma=\sqrt{\lambda^2-S^2/4}$, while the frequency of the eigenmode is $\Omega=-S/2$, where the minus sign derives from the difference between the coprecession frequency and the normal-mode frequency explained earlier.

In our second formulation of Eq.~\eqref{eq:second_formulation}, the pendulum parameters are different, with $\lambda_{v_1}^2=G_{v_1} D_{v_1}=K D_{v_1}(1-v_1 v_2) (v_2-v_1)$, and $S_{v_1}=\bJ_{v_1}\cdot \bP_{v_1}/D_{v_1}=(v_2-v_1)K+(1-v_1 v_2) D_0$. We can easily verify that the growth rate $\gamma=\sqrt{\lambda_{v_1}^2-S_{v_1}^2/4}$ is identical in this formulation, as it should, while the real part of the eigenfrequency is $\Omega'=\Omega-K v_2$, differing precisely by the corotation frequency that we introduced to obtain pendulum equations. Therefore, the instability properties are independent of the specific formulation. In the next section, we will show that the same properties are also recovered from linear stability~analysis.

The insight is new that a flavor pendulum can be formulated in many different ways and in particular, that any of the modes $\bP_w$ or $\bP_v$ can be taken as the pendulum radius vector. A broader discussion of this topic exceeds the scope of the present paper and will be presented elsewhere \cite{TFP}. The choice of variables that play the role of $\bG$, $\bR$, and $\bJ$ is much more flexible than had been traditionally appreciated.

\subsection{Dispersion relation}

Let us connect the properties of the two-beam pendulum with the instability that triggers it in the first place. This requires us to seek the unstable modes of the two-beam flavor-diagonal configuration, which we do in the conventional way by assuming $\Psi_v$ very small in Eq.~\eqref{eq:SW_eom_2} and $D_v$ fixed to their initial values. Assuming a solution $\Psi_v\propto e^{-i \Omega t}$, with $\Omega$ potentially complex, we immediately find the dispersion relation by requiring the consistency of $\Psi_0=\sum_v \Psi_v$ and $\Psi_1=\sum_v v \Psi_v$, where the sum extends over the two modes. This calculation is done in most papers on the dispersion relation of a neutrino plasma (see, e.g., Ref.~\cite{Fiorillo:2024uki} for its explicit form for axially symmetric systems), leading to
\begin{equation}
    (1-I_0)(1+I_2)+I_1^2=0,
    \quad\text{where}\quad
    I_n=\sum_v\frac{D_v v^n}{\Omega-kv}.
\end{equation}
Using the recursion relation
$I_n=(\omega I_{n-1}-D_{n-1})/k$
we can reduce the dispersion relation to the equivalent form
\begin{equation}\label{eq:modified_dispersion_relation}
    I_1=\frac{K\Omega}{kK-\omega\Omega}.
\end{equation}
For the two-beam setup, this is
\begin{equation}
    \frac{D_{v_1}v_1}{\omega-v_1 k}+\frac{D_{v_2}v_2}{\omega-v_2 k}=\frac{K\Omega}{kK-\omega \Omega}
\end{equation}
with the solution
\begin{eqnarray}\label{eq:general_solution_dispersion_twobeam}
    \Omega&=&\frac{1}{2}\bigl[K(v_1+v_2)-D_0(1-v_1 v_2)\bigr]
    \nonumber\\
    &&{}\pm\frac{1}{2} \sqrt{[K(v_1+v_2)-D_0(1-v_1 v_2)]^2-4[K^2 v_1 v_2-K D_1(1-v_1v_2)]}\,.
\end{eqnarray}
An instability ensues when the argument of the square root becomes negative. We can also check directly that the resulting growth rate coincides with the one predicted earlier from the pendulum parameters.

\subsection{Weakly unstable two-beam system}

The case of weak instability has particular relevance, especially in anticipation of Paper~II, where we will show that for weak instabilities the pendulum solution survives regardless of whether the system is two-beam or continuous. A weak instability emerges when one of the two polarization vectors, say $D_{v_1}$, has just become negative, while the other one $D_{v_2}$ is positive and large by comparison; this is a configuration with a very weak angular crossing and therefore barely unstable~\cite{Morinaga:2021vmc,Fiorillo:2024bzm}. So we can treat $D_{v_1}$ as a small perturbation in the square root in Eq.~\eqref{eq:general_solution_dispersion_twobeam}, and obtain the growth rate in the form
\begin{equation}
    \gamma^2\simeq-\frac{1}{4}\left\{[K(v_1-v_2)-D_{v_2}(1-v_1 v_2)]^2+2(1-v_1 v_2) D_{v_1} [K(v_1-v_2)+D_{v_2}(1-v_1 v_2)]\right\}.
\end{equation}
For very small $D_{v_1}$, the range of wavenumbers $K_-<K<K_+$ in which an instability can emerge is very small
\begin{equation}
    K_\pm\simeq \frac{(D_{v_2}-D_{v_1})(1-v_1 v_2)}{v_1-v_2}\pm \frac{2(1-v_1 v_2)}{v_1-v_2}\sqrt{-D_{v_1}D_{v_2}}.
\end{equation}
This confirms that an instability can only ensue if $D_{v_1}D_{v_2}<0$ and shows that the first unstable wavenumber to emerge is 
\begin{equation}\label{eq:firstK}
    K\simeq \frac{D_{v_2}(1-v_1 v_2)}{v_1-v_2}.
\end{equation}
From Eq.~\eqref{eq:general_solution_dispersion_twobeam}, the corresponding real part of the eigenfrequency for the first unstable mode and its growth rate are
\begin{equation}
    \Omega_R=\frac{D_{v_2}v_2(1-v_1 v_2)}{v_1-v_2}
    \quad\text{and}\quad
    \gamma=\sqrt{-D_{v_1}D_{v_2}}\,(1-v_1 v_2).
\end{equation}
The shifted wavenumber is $k=K+D_1=D_2(1-v_2^2)/(v_1-v_2)$, while the shifted frequency is $\omega=v_1 k$. The reason we emphasize these relations is that they highlight, once more, and even in a setup with discrete beams, that the first emergence of the instability always occurs through a resonance with the ``flipped'' beam. The phase velocity of the wave $\omega/k$ is indeed equal to the velocity of the beam with the small negative lepton number.

We can also determine the spin of the pendulum $S_{v_1}=\bJ_{v_1}\cdot \bR_{v_1}/|\bR_{v_1}|$ associated with the weak beam. Using Eq.~\eqref{eq:second_formulation} for the initially unstable configuration with all vectors aligned with the $z$ axis, we see immediately that
\begin{equation}
    S_{v_1}\simeq(v_2-v_1)K+(1-v_1 v_2)D_{v_2},
\end{equation}
where we have neglected $D_{v_1}\ll D_{v_2}$ in the last term. If we replace Eq.~\eqref{eq:firstK} for the value of $K$ associated with the first instability, the pendulum spin $S_{v_1}$ naturally vanishes.

The vanishing of the spin associated with the weak-beam pendulum has deeper significance. The instability itself has the character of a resonance, with the neutrinos moving in phase of the wave feeling the maximum effect and therefore flipping entirely in flavor under the action of the instability. Therefore, resonant neutrinos must behave as a plane pendulum, so that they can flip completely. In Paper~II, we will find that this conclusion extends beyond the exact pendulum discussed here; even for a continuous angular distribution, the resonant neutrinos behave approximately as a plane pendulum for a weak instability.

For a discrete set of beams, our conclusion about the weak instability resonating with one of the bins is completely general. A weakly unstable configuration is first reached by having all velocity modes with a positive lepton number difference $D_{v}$, except for one beam with a small negative $D_{v_i}$. From the structure of Eq.~\eqref{eq:modified_dispersion_relation}, we conclude that the corresponding contribution to the dispersion relation is initially small and negligible for all modes, which therefore cannot be qualitatively affected and turn unstable, \textit{except} for the modes with $\omega-v_i k$ very small. For these modes, the numerator and denominator are both small. In the limit $D_{v_i}\to 0$, while still being negative, this is the only mode that can be affected and turned unstable by the beam with flipped lepton number. In Paper~II, this observation will play a crucial role, since it will establish the connection between the two-beam exact pendulum, and the more general, albeit approximate, pendular behavior that characterizes the nonlinear saturation of all weak fast flavor instabilities.

\subsection{Continuous ensemble}

For a continuous $\bP_v$ system, or its numerical representation of many discrete beams, there does not usually exist a generalization of the two-beam pendulum, in contrast to the slow and fast spin systems. First, this becomes evident by numerical examples. Second, an analytic construction of an exact solution for the continuous system of the form Eq.~\eqref{eq:Lorentzian_general} has no discernible foundation because of the absence of Lax vectors. Starting from $\bK$ and two discrete modes, one cannot construct an ``interpolated'' system that would extend the two-beam pendulum to a continuous system.

In Paper~II, we will focus precisely on a continuous system, that produces additional surprises. However, from the practical perspective, we will find that for a weak instability, which occurs in a narrow resonant $v$-range in the region of the flipped population, this narrow range will once more act like a pendulum and shows the characteristic Lorentzian $v$-distribution.

\section{Discussion and summary}

\label{sec:discussion}

Motivated by the recent discovery of a new nonlinear solution of the fast flavor system~\cite{Liu:2025muc}, we have (i)~developed a deeper understanding of the origin and structure of this solution and (ii)~developed a taxonomy of the known solutions to illuminate how the new case fits into the grander picture of exact solutions of the nonlinear equations. We have found a very simple derivation of the new class of solutions based on the insight that a single-$K$ field of flavor coherence, together with homogeneous flavor occupation numbers, form a closed system of equations, without exciting other $K$-modes. This closed system is equivalent to a mechanical ensemble of interacting spins, in full analogy to the slow and fast homogeneous~systems.

All of these systems show coprecession solutions. In the slow and fast cases, these can be boosted to become nonlinear flavor waves with harmonic space and time variation and constant occupation numbers, in our equations represented by the variable $D_v$. For the SW system, the coprecession solution already is such a nonlinear flavor wave. In other words, on the level of the coprecession solutions, the SW system is nothing but another way to look at nonlinear flavor waves from a boosted frame. It is not a new solution.

Of greater potential practical interest are configurations that connect by an instability to a system that is initially in a nearly pure flavor state without large flavor coherence. In this case, the SW system does yield a new solution, an unstable system with harmonic spatial variation and a run-away temporal structure. However, there are significant limitations. An exact pendulum-like solution exists only for two velocity beams, which cannot be extended to a continuum of modes, in contrast to the slow and fast homogeneous systems. The underlying reason is the lack of Gaudin invariants for the SW system, in contrast to the previous cases. 

In Paper~II of this series \cite{Fiorillo:2026vfo}, we will examine the SW system with a continuous angle distribution, with a particular focus on weak instabilities, which in any realistic system will be the first to appear. In this case, when only a narrow range of $v$-modes resides within the resonance region and participates in flavor conversion, this narrow range will once more behave as a flavor pendulum, over the entire range of participating resonant modes.

The SW system is the third mechanical flavor spin system for which instabilities naturally connect, in the nonlinear regime, to pendulum-like solutions, although here only for a two-beam case. The pendulum dynamics turns out to be much more generic for interacting classical spin systems than had been previously realized. We will explore the general theory of the flavor pendulum elsewhere~\cite{TFP}.

\section*{Acknowledgments}

We thank Lucas Johns, Dwaipayan Mukherjee, G\"unter Sigl, Irene Tamborra, and Cristina Volpe for thoughtful comments on an early draft. DFGF is supported by the Alexander von Humboldt Foundation (Germany). GGR acknowledges partial support by the German Research Foundation (DFG) through the Collaborative Research Centre ``Neutrinos and Dark Matter in Astro- and Particle Physics (NDM),'' Grant SFB--1258--283604770, and under Germany’s Excellence Strategy through the Cluster of Excellence ORIGINS EXC--2094--390783311.

\bibliographystyle{JHEP}
\bibliography{References}

\end{document}